\documentclass[aps,pra,eqsecnum,twocolumn,showpacs,superscriptaddress]{revtex4}

\usepackage{graphicx}
\usepackage{amsmath}
\usepackage{amssymb}
\usepackage{bm}

\newcommand{\tr}{\mathop{\text{Tr}}\nolimits}
\newcommand{\ket}[1]{|{#1}\rangle}
\newcommand{\bra}[1]{\langle{#1}|}

\newcommand{\beq}{\begin{equation}}
\newcommand{\eeq}{\end{equation}}
\newcommand{\barr}{\begin{eqnarray}}
\newcommand{\earr}{\end{eqnarray}}

\newcommand{\andy}[1]{}

\begin{document}

\title{Robust gates for holonomic quantum computation}

\author{Giuseppe Florio}
\affiliation{Dipartimento di Fisica, Universit\`a di Bari,
        I-70126  Bari, Italy}
\affiliation{INFN, Sezione di Bari, I-70126 Bari, Italy}
\author{Paolo Facchi}
\affiliation{Dipartimento di Matematica, Universit\`a di Bari,
        I-70125  Bari, Italy}
\affiliation{INFN, Sezione di Bari, I-70126 Bari, Italy}
\author{Rosario Fazio}
\affiliation{NEST CNR-INFM $\&$ Scuola Normale Superiore, Piazza
        dei Cavalieri 7, 56126 Pisa, Italy}
\affiliation{International School for Advanced Studies (SISSA)
        via  Beirut 2-4,  I-34014 Trieste, ITALY}
\author{Vittorio Giovannetti}
\affiliation{NEST CNR-INFM $\&$ Scuola Normale Superiore, Piazza
        dei Cavalieri 7, 56126 Pisa, Italy}
\author{Saverio Pascazio}
\affiliation{Dipartimento di Fisica, Universit\`a di Bari,
        I-70126  Bari, Italy}
\affiliation{INFN, Sezione di Bari, I-70126 Bari, Italy}

\begin{abstract}
Non Abelian geometric phases are attracting increasing interest
because of possible experimental application in quantum computation.
We study the effects of the environment (modeled as an ensemble of
harmonic oscillators) on a holonomic transformation and write the
corresponding master equation. The solution is analytically and
numerically investigated and the behavior of the fidelity analyzed:
fidelity revivals are observed and an optimal finite operation time
is determined at which the gate is most robust against noise.

\end{abstract}

\date{\today}


\maketitle

\section{Introduction}
\label{sec: Introduction}

One of the possible alternatives to quantum information
processing~\cite{nielsen,casati}, which is attracting increasing
interest, is based on geometric interferometry~\cite{shapere,bohm}.
In this case, the transformations needed to implement the quantum
gates are realized by making the Hamiltonian of the quantum computer
dependent on a set of controlling parameters which describe suitable
closed loops in an associated parameter space. In the adiabatic
limit the dynamical contribution to the evolution can be factorized
and the quantum gate only depends on the topological structure of
the manifold. This should be contrasted with the dynamical approach
to quantum computation, where the desired phase factors in the
quantum gates are of dynamical origin.  Geometric quantum
computation has been formulated using both Abelian~\cite{jones} and
non-Abelian~\cite{zanardi} holonomies. In Ref.\ \cite{jones} the
first experimental demonstration of geometric quantum gates using
nuclear magnetic resonance was presented. Since the appearance of
the original proposals, several studies addressed the implementation
with quantum optical~\cite{duan}, superconducting~\cite{falci,faoro}
and semiconducting systems~\cite{solinas1}.

As is well known, decoherence is detrimental for quantum
computation. Despite the large body of knowledge accumulated to
study decoherence in open quantum systems~\cite{giulini}, the study
of geometric phases in the presence of decoherence and dissipation
has started only recently, although with a few exceptions, and was
certainly prompted by the interest in quantum computation. Together
with many common  features with the theory of open quantum systems,
the analysis of decoherence in geometric interferometry rises
several distinct issues that are of interest both as fundamental
questions in quantum mechanics and in quantum computation. The
adiabatic evolution, for example, cannot occur arbitrarily slow, as
decoherence would destroy any interference. This implies that the
decoherence processes should be analyzed in close connection with
non-adiabatic corrections, a question which is not typically present
in the non-unitary dynamical evolution of open systems. Moreover,
the period of the evolution fixes a new time scale, compared to
which the different components of the bath will act differently.
Finally, in the non-Abelian case the coupling to the environment may
(partially) lift some degeneracy and therefore modify the holonomy
itself. These are only a few examples of questions which emerge when
one wants to study geometric phases in open systems.

Most of the attention on the properties of geometric phases in the
presence of coupling to an external bath has focused on the
Abelian~\cite{ellinas,gamliel,
dechiara,whitney1,whitney2,carollo,sarandy1,ericsson,marzlin,kamleitner}
case. There are however a few important exceptions where non-Abelian
holonomies in open systems have been investigated as
well~\cite{solinas2,fuentes,sarandy2,parodi}. Solinas {\em et
al.}~\cite{solinas2} studied the influence of parametric noise on
the scheme for holonomic quantum computation discussed in Ref.\
\cite{solinas1}. Parodi~\emph{et al.}~\cite{parodi} analyzed the
effects of different spectral densities of a quantum thermal bath on
the efficiency of this scheme. The quantum jump approach was applied
by Fuentes-Guridi \emph{et al.}\ \cite{fuentes} in order to
understand under which circumstances holonomic quantum computation
is robust against decoherence. Very recently Sarandy and
Lidar~\cite{sarandy2} analyzed non-Abelian holonomies for open
systems, starting from an analysis of the master equation in
Lindblad form for the reduced density matrix.

The aim of this work is to study the non-adiabatic dynamics and
the effects of quantum noise on the setup proposed by Duan, Cirac
and Zoller~\cite{duan}. We will present  a  class of 1-qubit
holonomic quantum gates (which includes the NOT gate) that are
intrinsically robust against any type of noise.  The only
requirement is that the noise be sufficiently small, so that a
master equation can be written. The above mentioned robustness is
a consequence of a peculiar property of this class of gates,
namely the possibility to realize in the noiseless case a
\emph{perfect} gate transformation (i.e.\ with fidelity one) in a
\emph{finite} time. In particular this class exhibits fidelity
revivals, that consist in an infinite number of (almost periodic)
time values at which the fidelity reaches unity. The first revival
is the \emph{optimal operational time} for a nonadiabatic gate in
presence of noise, because it represents the point with the
highest fidelity among all revivals, not to mention the
corresponding adiabatic gate, whose fidelity is far lower. In this
respect our analysis consists in a generalization to non-adiabatic
holonomic quantum computation. Non-adiabatic holonomies have been
discussed by Anandan~\cite{anandan88}, generalizing to the
non-Abelian case the work of Aharonov and Anandan~\cite{aharonov}.
Very recently, the use of non-adiabatic phases has been discussed
in the framework of geometric computation as a way to further
protect the computer from decoherence~\cite{blais}. In the same
spirit we discuss this possibility in the non-Abelian case. We
derive a master equation for the reduced density matrix of the
system in the presence of a bath in the weak coupling
approximation. This equation is numerically and analytically
solved, displaying revivals of the fidelity and the existence of
the afore-mentioned ``optimal'' finite operation time at which the
detrimental effects of decoherence are minimized.

This paper is organized as follows. In Section \ref{sec:
Non-Abelian Geometric Phases} we briefly review the concept of
holonomy in the absence of the environment and introduce notation.
In Section
\ref{sec: Free Ideal Evolution} we focus our attention on the
specific physical system discussed in \cite{duan} and study the
role of non-adiabatic effects, which turn out to be important when
discussing the realistic case of a finite time evolution in the
presence of the environment, which we present in Section~\ref{sec:
Noise and Master Equation}. We derive here a general master
equation for time dependent Hamiltonians and, after specializing
it to our case, we numerically solve it in Section \ref{sec:
results}. The fidelity of the operations in the presence of noise
will be discussed in detail, using the ideal case as a reference.
Conclusions and further perspectives are presented in
Section~\ref{sec: conclusions}.

\section{Holonomies}
\label{sec: Non-Abelian Geometric Phases}

Let us introduce notation. Suppose that a system, governed by a non
degenerate Hamiltonian that depends on time through a set of
parameters, evolves adiabatically, covering a closed loop in the
parameter space. Berry~\cite{berry} discovered that at the end of
the evolution the final state exhibits, in addition to the dynamical
phase, also a geometric phase, whose structure depends only on the
topological properties of the manifold on which the system has
evolved.

The situation changes if the Hamiltonian possesses some
degeneracies. In this case a loop in the parameter space realizes
more complex geometric transformations \cite{wilczeck}. Let us
assume that  the system eigenspaces (indexed by $m$) are
degenerate  and denote by $|m_k(t)\rangle$ their set of
instantaneous eigenstates (the degeneracy index $k$ ranging from
$1$ to $N_m$). The instantaneous eigenstates form an orthonormal
basis
\beq
    \langle m'_{k'}(t)|m_k(t)\rangle=\delta_{mm'}\,\delta_{kk'} .
\label{eq: norm condition}
\eeq
The time evolution of the quantum system is governed by the
Schr\"{o}diger equation
\beq
    i\frac{d}{d t}|\psi(t)\rangle=H(t)|\psi(t)\rangle ,
\label{eq: Schroedinger equation}
\eeq
with $H$ depending on $t$ through a set of
parameters~$x^{\mu}(t)$. Having in mind quantum computation
applications we will suppose that the family of Hamiltonians
$H(x(t))$ is iso-degenerate, i.e.\ that the dimensions of its
eigenspaces do not depend on the parameters and that its
eigenprojections $P_m(x(t))$ have a smooth dependence on $t$ (at
least twice continuously differentiable). In particular, this
implies that there is no level crossing between different
eigenspaces. At each time $t$, $H(t)$ can be decomposed by using
its instantaneous eigenprojections
$P_m(t)=\sum_k|m_k(t)\rangle\langle m_{k}(t)|$:
\beq
    H(t)=\sum_{m}\epsilon_m(t) P_m(t) .
\label{eq: decomp H}
\eeq
We define the operator $R$, whose action transports every
eigenprojection from $t_0$ to $t$,
\beq
R(t,t_0) P_m(t_0) = P_m(t) R(t,t_0).
\label{eq:intertwin}
\eeq
Its generator is hermitian ($D(t,t_0)=D^\dag(t,t_0)$) and reads
\barr
& & D(t,t_0)= -i R(t,t_0)^\dagger \frac{\partial}{\partial
t}R(t,t_0).
\label{eq:defD}
\earr
In the interaction picture defined by the operator $R$ we have
\beq
\label{eq: }
\tilde H(t,t_0)=R^{\dag}(t,t_0)H(t)R(t,t_0)=\sum_{m} \epsilon_m
(t)P_m(t_0).
\eeq
and the evolution operator can be written as
\beq\label{eq: decomp RU}
U(t,t_0)= R(t,t_0) \textbf{T}\exp\left\{-i\int_{t_0}^{t}(\tilde
H(s,t_0)+D(s,t_0))ds\right\}
\eeq
$\textbf{T}$ being the chronological product. In the adiabatic
limit the evolution of the state remains confined in the
degenerate eigenspaces. The above evolution operator becomes
block-diagonal and, in the case of cyclic evolution
($P_m(t)=P_m(t_0)$), reads:
\begin{equation}\label{eq:Uadiab}
U(t,t_0)\sim \sum_m P_m(t_0)e^{-i  \int_{t_0}^{t} \epsilon_m(s)
ds} U_{\mathrm{ad}}^m P_m(t_0),
\end{equation}
where the geometric evolution
\beq
    U_{\mathrm{ad}}^m =\textbf{P}\exp\left\{-\oint_C A^m(x)\right\}
\label{eq: geometric evolution}
\eeq
is given by a path ordered integral of the adiabatic connection
$A^m(x)=\sum_\mu A^m_\mu dx^\mu$, with
\barr
A^m_\mu(x(t)) &=& P_m(x(t_0)) R^{\dag}(x(t),x(t_0))
\nonumber\\
& &\times \frac{\partial}{\partial
x^\mu}R(x(t),x(t_0))P_m(x(t_0)).
\label{eq: adiabatic connection}
\earr
The holonomy thus obtained is the fundamental ingredient for
realizing complex geometric transformations. In the following
section we will give an explicit example.

In the following we will also take into account nonadiabatic
effects under the simplifying assumption that the eigenvalues
$\epsilon_m$ are time-independent and the connection
(\ref{eq:defD}) is piecewise constant.
Then the operator $D(t,t_0)=D(t_0,t_0)$ does not depend on time
during the evolution $\,\,\forall s\,\in\,[t,t_0]$.
Moreover, $\tilde H(t,t_0)=H(t_0)$ in (\ref{eq: }) and
Eq.~(\ref{eq: decomp RU}) reduces to the useful expression
\beq\label{eq: U with D indip t}
U(t,t_0)=e^{i(t-t_0)D(t_0,t_0)}\,e^{-i(t-t_0)(H(t_0)+D(t_0,t_0))},
\eeq
which, in the adiabatic limit, becomes
\beq\label{eq: Uadiab with D indip t}
U(t,t_0)\sim \sum_m P_m(t)e^{-i  (t-t_0)
\left[H(t_0)+A^m(t_0)\right]} P_m(t_0).
\eeq

We will see that for a large class of gates it is possible to
evaluate exactly the time evolution, including all non-adiabatic
effects. The analysis of the evolution operator will enable us to
find an optimal working point where the gate is robust against
noise. This optimal time is related to revivals of fidelity, i.e.\
(finite) values of time at which the fidelity goes back to 1.

\section{Free Ideal Evolution for a Tripod System}
\label{sec: Free Ideal Evolution}

\subsection{Preliminaries}

\begin{figure}
\includegraphics[width=0.4 \textwidth]{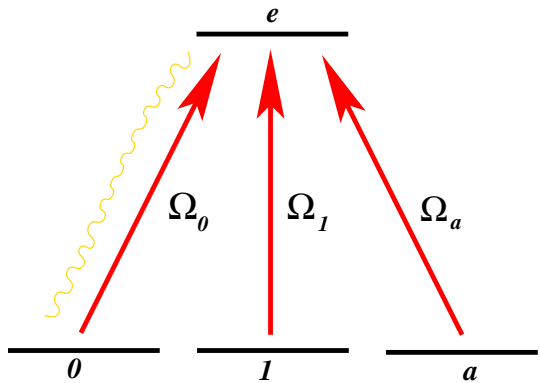}
\caption{(Color online) Scheme of a tripod system: 0 and 1 are computational
levels, while $a$ is an ancilla state used for the intermediate
steps of the transformation. The three degenerate levels are
connected with an upper level $e$ by time dependent Rabi
frequencies $\Omega_j(t)$. We also show the noise introduced in
our analysis, that induces additional transitions between $0$ and
$e$.}
\label{fig: levels}
\end{figure}

We consider the system introduced in~\cite{duan} for holonomic
quantum computation: see Fig.\ \ref{fig: levels}, where three
degenerate levels are connected with a fourth one by Rabi
oscillations. The adiabatic evolution of this system was analyzed
in several papers for different experimental implementations
\cite{duan,faoro,solinas1,unanyan}. Here we review the ideal
noiseless case, taking into account also non-adiabatic effects
that are important in the presence of decoherence, when the loop
cannot be completed in an arbitrarily long time. At time $t=0$ the
logical states $0$ and $1$ are encoded respectively in $|0\rangle$
and $|1\rangle$, while $|a\rangle$ is an ancilla state used as
``buffer'' during the evolution.
The Hamiltonian of the system reads
\beq
    H(t)=|e\rangle(\Omega_0(t)\langle 0|+\Omega_1(t)\langle
    1|+\Omega_a(t)\langle a|)+ \mathrm{H.c.}
\label{eq: hamiltonian S}
\eeq
where $\Omega_j(t)$ represent the time dependent Rabi frequencies
of the transitions. The loop in the parameter space is obtained by
varying $\Omega_j(t)$  ($j=0,1,a$). In our calculations we
consider $\Omega_j(t)\in \mathbb{R}, \forall t$. The eigenvalues
of the system are
\beq
    \{0,\pm\sqrt{\Omega_0(t)^2+\Omega_1(t)^2+\Omega_a(t)^2}=\pm
    \Omega\} ,
\label{eq: eigenvalues}
\eeq
where $0$ is 2-fold degenerate, corresponding to a 2-dimensional
(computational) eigenspace, and $\Omega$ is kept constant.
Therefore, the parameter space is the 2-sphere of radius $\Omega$,
given by $\{\Omega_j\in\mathbb{R}|\sum_j\Omega_j^2=\Omega^2\}$.
Introducing the parametrization
\begin{equation}
\Omega_1=\Omega\,\sin{\vartheta}\,\cos{\varphi}, \;
\Omega_0=\Omega\,\sin{\vartheta}\,\sin{\varphi}, \;
\Omega_a=\Omega\,\cos{\vartheta} ,
\label{eq: param trans}
\eeq
the eigenstates take the form
\begin{widetext}
\begin{eqnarray}
    |D_0(t)\rangle&=&
    \cos{\varphi}\,|0\rangle-\sin{\varphi}\,|1\rangle,  \nonumber\\
    |D_1(t)\rangle&=&\cos{\vartheta}\, \sin{\varphi}|0\rangle+\cos{\vartheta}\,
    \cos{\varphi}|1\rangle-\sin{\vartheta}|a\rangle, \label{eq: eigenstates in param}\\
    |D_\pm(t)\rangle&=&\big(\pm|e\rangle+\sin{\vartheta}\, \sin{\varphi}|0\rangle+
    \sin{\vartheta}\,
    \cos{\varphi}|1\rangle+\cos{\vartheta}|a\rangle\big)/\big(\sqrt{2}\big).\nonumber
\earr
\end{widetext}
The computational space (belonging to the degenerate eigenvalue
$0$) is
\beq
    C_S=\text{Span}\{|D_0(t)\rangle,|D_1(t)\rangle\} ,
\label{eq: computational space}
\eeq
while $|D_\pm(t)\rangle$ are the bright eigenstates belonging to
$\pm\Omega$.

Applying the definition given in Section \ref{sec: Non-Abelian
Geometric Phases}, the elements of the adiabatic connection
(\ref{eq: adiabatic connection}) for the eigenspace $m=0$ are
\begin{equation}
A_{\vartheta}= 0, \qquad
A_{\varphi}= i \sigma_y \cos\vartheta ,
\label{eq: connection adiab}
\eeq
where
$\sigma_y=-i(\ket{D_0(t_0)}\bra{D_1(t_0)}-\ket{D_1(t_0)}\bra{D_0(t_0)})$.
The holonomy (\ref{eq: geometric evolution})  for a closed loop on
a sphere for the computational space reads
\beq\label{eq: ev op adiab}
U_{\mathrm{ad}}=\textbf{P}e^{-\oint_C A_\varphi d\varphi} = e^{-i
\sigma_y \oint_C \cos\vartheta d\varphi}=\exp\left(i \sigma_y
\,\mathcal{\omega}\right) ,
\eeq
where $\omega$ is the solid angle enclosed by the loop in the
parameter space. As an explicit example (that will be considered
in the following), let $\omega=\pi/2$, and obtain
\beq
U_{\pi/2}=\exp(i \sigma_y \,\pi/2)= i \sigma_y=\left(\begin{array}{cc}0&1\\
    -1&0\\
    \end{array}\right)
\label{eq: NOT oper in the adiab lim}
\eeq
(in the basis $\{|D_0(t_0)\rangle,|D_1(t_0)\rangle\}$), that
represents a NOT transformation (up to a phase for the state
$|D_0\rangle$).

Following the discussion in the previous Section we discuss the
non-adiabatic corrections to this system. In order to use
Eq.~(\ref{eq: U with D indip t}), we will consider the loop shown
in Fig. \ref{fig: NotPath}, obtained by going from the pole to the
equator, spanning then a $\pi/2$ angle and finally going back to
the pole. The solid angle enclosed in this loop is equal to
$\pi/2$; in the adiabatic limit (when the product $\Omega\,\tau$
goes to infinity, $\tau$ being the total time of the cyclic
evolution and $\Omega$ the energy of the bright states) this path
yields a NOT gate as in Eq.~(\ref{eq: NOT oper in the adiab lim}).

\begin{figure}
\includegraphics[width=0.3 \textwidth]{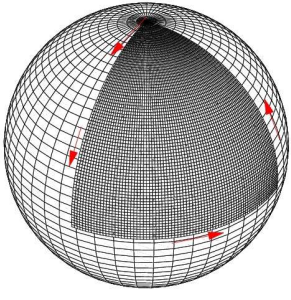}
\caption{(Color online) Path in parameter space for the realization of a NOT
gate. The solid angle spanned during the evolution is $\pi/2$.}
\label{fig: NotPath}
\end{figure}

The first step consists in constructing the operator $D$ from
Eq.~(\ref{eq: eigenstates in param}) and the definition
(\ref{eq:defD}); its matrix representation, written in the basis
$\{\ket{D_i(t_0)}\}_{i=0,1,+,- }$, is
\begin{widetext}
\beq
    D(t,t_0)=-i\left(\begin{array}{cccc}
    0&{\dot{\varphi}\,\cos{\vartheta}}&{\dot{\varphi}\,
    \sin{\vartheta}/\sqrt{2}}&{\dot{\varphi}\,\sin{\vartheta}/\sqrt{2}}\\
    {-\dot{\varphi}\,\cos{\vartheta}}&0&{\dot{\vartheta}/\sqrt{2}}&{\dot{\vartheta}/\sqrt{2}}\\
    {-\dot{\varphi}\,\sin{\vartheta}/\sqrt{2}}&{-\dot{\vartheta}/\sqrt{2}}&0&0\\
    {-\dot{\varphi}\,\sin{\vartheta}/\sqrt{2}}&{-\dot{\vartheta}/\sqrt{2}}&0&0\\
    \end{array}\right) .
\label{eq: connection}
\eeq
\end{widetext}

\subsection{Analytical results}

One can see from Eq.~(\ref{eq: connection}) that, as far as the
rate of change of the angles $\varphi$ and $\vartheta$ is constant
in each segment of the path, we can use Eq.~(\ref{eq: U with D
indip t}) to calculate the evolution operator along the path shown
in Fig.\ \ref{fig: NotPath}. The complete expression is explicitly
given in Appendix \ref{sec: Appendix A}.

A noteworthy feature of the exact expression is that it is
factorized in three terms. In the adiabatic limit it simplifies to
\begin{eqnarray}
U_{\pi/2}(\Omega\tau)&=&U_{3}(\Omega\tau_3)\,U_{2}(\Omega\tau_2)\,U_{1}(\Omega\tau_1)
\stackrel{\tau \Omega\rightarrow+\infty}{\longrightarrow}\nonumber\\
\nonumber\\
    U_{\pi/2}^{\rm{ad}}(\Omega\tau)&=&\left(\begin{array}{cccc}0&1&0&0\\
    -1&0&0&0\\
    0&0&e^{-i\tau\Omega}&0\\
    0&0&0&e^{+i\tau\Omega}\\
    \end{array}\right) ,
\label{eq: U analytic in adiab lim}
\end{eqnarray}
$\tau$ being the total evolution time needed for covering the loop
in the parameter space and $\tau_i=\alpha_i\tau$, with
$\sum_i\alpha_i=1$. This represents a NOT gate for the degenerate
subspace and yields (fast oscillating) dynamical phases for the
bright states.


\subsection{Fidelity revivals}

In order to understand how far the evolution operator is from the
ideal one, we use the fidelity, defined as
\begin{eqnarray}
F(\tau)&=&\tr\{\sigma_{\rm{ad}}(\tau)\sigma(\tau)\}\nonumber\\
&=& \tr\{U_{\pi/2}^{\rm{ad}}\sigma(0){U^{\rm{ad}\dag}_{\pi/2}}
\,U_{\pi/2}(\Omega\tau)\sigma(0)U^{\dag}_{\pi/2}(\Omega\tau)\} ,
\nonumber\\
\label{eq: def fidelity}
\end{eqnarray}
where $\sigma(0)$ is the density operator describing the initial
state, assumed to be pure, and $\sigma_{\rm{ad}}$ the
corresponding operator for the adiabatic ideal evolution. The mean
fidelity (averaged over a set of input states uniformly
distributed on the Bloch sphere) is plotted in Fig.\
\ref{fig:ideal} as a function of the adiabaticity parameter
$\Omega\tau$.
\begin{figure}
\includegraphics[width=0.4 \textwidth]{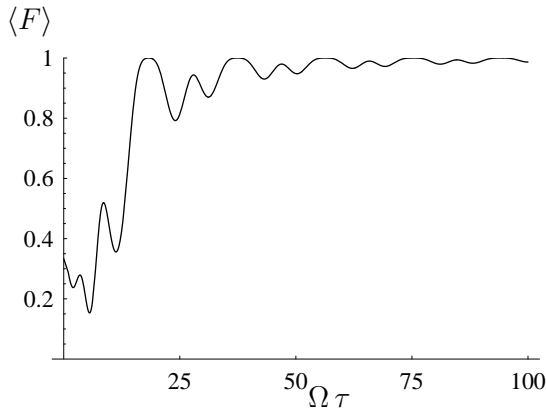}
\caption{Mean fidelity versus the cyclic time
$\Omega \tau$ (noiseless case). $\Omega$ is the energy gap between
the bright and dark states. $\tau$ is the time needed to cover the
loop shown in Fig.~\ref{fig: NotPath}. The average is performed
over a set of initial states uniformly distributed on the Bloch
sphere.}
\label{fig:ideal}
\end{figure}

$F(\tau)$ asymptotically approaches the value 1 (with some
oscillations), as expected (adiabatic limit). Interestingly, the
fidelity is exactly one for some \emph{finite} values of time,
$\tau=\tau^*_k$, that are clearly independent of the initial
state. In this case the NOT transformation is perfect, even though
one is far from the adiabatic regime. We now discuss this curious
feature that will turn out to be of interest in the search for an
optimal operation time at which the computation is most robust
against noise (see Section \ref{sec: results}).

In order to obtain a formula for the times $\tau^*_k$, consider
the operator
\begin{equation}\label{eq:Q}
 Q(\Omega\tau)=U^{\dag}_{\pi/2}(\Omega\tau)U_{\pi/2}^{\rm{ad}}(\Omega\tau)
\end{equation}
appearing in Eq.~(\ref{eq: def fidelity}). When the three arcs in
the loop in Fig.\ \ref{fig: NotPath} are covered in equal times, one
obtains a lengthy analytical expression for $Q$, not reproduced
here. In particular, the (1,1) element reads
\beq\label{eq: element (1,1) in Q}
Q_{11}(\Omega\tau)=\frac{4\,{\Omega }^2\,{\tau }^2 +
    9\,{\pi }^2\,\cos \left(\frac{\Omega\tau}{3}
    \sqrt{1 +\left(\frac{3\pi}{2\Omega \tau}\right)^2}\right)}{9\,{\pi}^2 +
    4\,{\Omega }^2\,{\tau }^2}
\eeq
and by equating the above expression to $1$ one gets
\beq\label{eq: optimal times for NOT}
\tau^*_k= \frac{3\pi}{2\Omega}\sqrt{16 k^2-1} \ , \qquad k\in
\mathbb{N}^* .
\eeq
In turn, this yields, by direct substitution in Eq.~(\ref{eq:Q}),
\beq\label{eq: Q in tau*}
Q(\Omega\tau^*_k)=\left(\begin{array}{cccc}1&0&0&0\\
0&1&0&0\\
0&0&e^{-i\Omega\tau^*_k}&0\\
0&0&0&e^{+i\Omega\tau^*_k}\\
\end{array}\right) .
\eeq
The first fidelity revival is obtained for $k=1$ and their
approximate frequency reads
\beq\label{eq: frequency of optimal points}
f(\tau,\Omega)=\frac{\Omega }{3}\sqrt{1
+\left(\frac{3\pi}{2\Omega\tau}\right)^2} .
\eeq
From this equation it is clear that the periodicity of the maxima in
Fig.\ \ref{fig:ideal} is only apparent (the frequency depends on
$\Omega\tau$). On the other hand, the second term in the square root
is very small, even for the first peak, making $f/\Omega$
approximately constant. The additional seemingly optimal points
appearing in Fig.\ \ref{fig:ideal} (at approximately double
frequency) are not solution of Eq.\ (\ref{eq: optimal times for
NOT}).

These revivals can be important for experimental applications: in
principle they would enable one to obtain a perfect NOT
transformation, without reaching the adiabatic regime. It is
important to notice that this result does not depend on the initial
state of the system but it is a feature of the chosen path: indeed
the operator $Q(\Omega\tau)$ does not contain any information about
the initial state.

Finally, we emphasize that
similar features (and in particular the presence of the revivals
in the non-adiabatic regime) hold for a large class of gates. For
transformations consisting in a loop which starts at the pole,
spans a segment on a geodesics (the equator) and goes back to the
pole enclosing a solid angle $\omega=\pi/2n$ ($n\in \mathbb{N}^*$)
there is a straightforward generalization of Eq.~(\ref{eq: optimal
times for NOT}):
\beq \label{eq: generalization of optimal times}
\tau^*_k(n)= \frac{(2n+1)\pi}{2n\Omega}\sqrt{16 k^2 n^2-1}.
\eeq
This expression is valid provided that the loop is covered at a
constant angular speed:
\beq
\dot{\vartheta}_{\rm segment 1}=\dot{\varphi}_{\rm segment
2}=\dot{\vartheta}_{\rm segment 3}=\mathrm{const}.
\eeq
Reversing the orientation of the loops leads to identical results.
These general observations can be of interest for experimental
applications.

\section{Noise and Master Equation}\label{sec: Noise and Master Equation}

A physical system is never completely isolated from its environment
and, in order to take into account the effects produced by the
latter, one analyzes the dynamics in terms of a master equation. In
the usual approach to this problem one assumes that the Hamiltonian
of the system is time independent (see for instance
\cite{gardiner}). For time dependent Hamiltonians a slightly
different approach is needed. Rigorous mathematical results were
derived by Davies and Spohn \cite{davies}. We summarize here the
main conclusions, providing for completeness a physical derivation
in Appendix \ref{sec: Appendix B}, where emphasis is put on the
physical meaning of the analysis in the context of adiabaticity and
holonomic quantum computation.

We consider a general Liouville operator with a time dependent
system Liouvillian
\beq\label{eq: liouvillianA}
\mathcal{L}(t)=\mathcal{L}_0(t)+\mathcal{L}_{SB}
=\mathcal{L}_S(t)\otimes1+1\otimes\mathcal{L}_{B}+\mathcal{L}_{SB}
.
\eeq
The evolution of density operator $\varrho(t)$, describing the
system and the environment, is governed by the von
Neumann-Liouville equation
\beq\label{eq: def eq liouvilleA}
\dot{\varrho}(t)=\mathcal{L}(t)\,\varrho(t) .
\eeq
We assume that there are no initial correlations between system
(whose density matrix is $\sigma$) and bath (whose density matrix
is $\sigma_B$) and that the latter is in equilibrium (e.g.\ in a
thermal state)
\beq\label{eq: factorization+bath equilA}
\varrho(0)=\sigma(0)\otimes\sigma_B, \qquad
\mathcal{L}_{B}\sigma_B=0 .
\eeq
The key hypothesis in the derivation of a master equation is that
the typical timescale of the evolution is much slower than the
timescales characterizing the bath. The additional hypothesis in
our case is that the timescale related to the rate of change of
the system Hamiltonian is the slowest timescale of our problem,
due to the adiabaticity of the evolution. In other words, compared
to the bath correlation time, the evolution of $\mathcal{L}_S$ is
always ``adiabatic." This is assured by the condition
\beq\label{eq: condition on LSA}
\tau_c\Delta\ll1 ,
\eeq
where $\tau_c$ is the correlation time of the bath and the energy
gap, $\Delta=
{\text{min}}\left|\epsilon_n(t)-\epsilon_m(t)\right|$, characterizes
the rate of change of $\mathcal{L}_S$. Under these conditions one
gets (Appendix \ref{sec: Appendix B})
\begin{eqnarray}\label{eq:MEred}
\dot{\sigma}(t)
=[\mathcal{L}_{S}(t)+\Gamma(t)]\sigma(t) ,
\end{eqnarray}
where $\sigma(t)= \tr_B\left\{\varrho(t)\right\}$ is the system
density matrix and
\begin{eqnarray}
\Gamma(t)&=& \sum_{\omega}Q_{\omega}(t)\int_{-\infty}^{0}du
\tr_B\left\{\mathcal{L}_{SB}\exp[-\mathcal{L}_0(t)\,u]\right.\nonumber\\
& & \times \left.\mathcal{L}_{SB}
\exp[\mathcal{L}_0(t)\,u]\sigma_B\right\}Q_{\omega}(t) ,
\label{eq: corr functA}
\end{eqnarray}
$Q_{\omega}(t)$ being the instantaneous eigenprojections of
$\mathcal{L}_S$,
\begin{eqnarray}
\label{eq: Liouvillian projectorsA}
\mathcal{L}_S(t) &=& i\sum_{\omega}\omega(t)Q_{\omega}(t), \nonumber\\
\sum_{\omega}Q_{\omega}(t)&=&1,\quad
Q_{\omega}(t)Q_{\omega'}(t)=\delta_{\omega \omega'}Q_{\omega}(t).
\quad
\end{eqnarray}
Equation (\ref{eq:MEred}) is the same master equation one would
obtain by considering $\mathcal{L}_S(t)$ ``frozen" at time $t$ and
evaluating the decay rates and the frequency shifts at the
instantaneous eigenfrequencies
$\omega(t)=\epsilon_m(t)-\epsilon_n(t)$ of the system Liouvillian.

We now turn our attention to the physical system described in
Sec.\ \ref{sec: Free Ideal Evolution}. In terms of the total
Hamiltonian,
\beq\label{eq: def liouvillian}
\mathcal{L}(t) \rho =-i[H_T(t),\rho] .
\eeq
For simplicity we consider an environment affecting only the
transitions between levels $|0\rangle$ and $|e\rangle$; this is
enough for our purposes. The total Hamiltonian is
\beq\label{eq: general hamiltonian with noise}
H_T(t)=H(t)+H_B+ \lambda H_{SB},
\eeq
where $\lambda$ is a dimensionless scaling factor introduced for
later convenience and representing the strength of the noise, and
$H(t)$ is the system Hamiltonian (\ref{eq: hamiltonian S}). The
bath is an ensemble of quantum harmonic oscillators,
\beq\label{eq: hamiltonian bath}
H_B=\sum_k  \omega_k{a_k}^\dag a_k ,
\eeq
with $\omega_k$ the frequency of the $k$-th mode. The interaction
Hamiltonian is
\beq\label{eq: hamiltonian interaction}
H_{SB}=\sum_k \gamma_k (|0\rangle\langle e| +|e\rangle\langle
0|)\otimes({a_k}^\dag+a_k) ,
\eeq
where $\gamma_k$ is the coupling constant between the system and
the $k$-th mode of the bath. By using Eq.~(\ref{eq: eigenstates
in param}) we can write
\begin{eqnarray}
& &|e\rangle\langle 0|=
\frac{\cos{\varphi}}{\sqrt{2}}(|D_+(t)\rangle-|D_-(t)\rangle)\langle
D_0(t)|\nonumber\\
& & + \frac{\sin{\varphi}\cos{\vartheta}}{\sqrt{2}}
(|D_+(t)\rangle-|D_-(t)\rangle)\langle D_1(t)|\nonumber\\
& & + \frac{\sin{\varphi}\sin{\vartheta}}{2}
(|D_+(t)\rangle-|D_-(t)\rangle)\left(\langle
D_+(t)|+\langle D_-(t)|\right). \nonumber\\
\label{eq: decomposition jump oper}
\end{eqnarray}
In the interaction picture generated by the operator $R$ defined in
(\ref{eq:intertwin}), the density operator reads
\beq\label{eq: dens matr in R representation}
\rho_R(t)=R^\dag\sigma(t)R .
\eeq
By taking the time derivative of Eq.~(\ref{eq: dens matr in R
representation}) we obtain
\beq\label{eq: dens matr in R representation2}
\dot{\sigma}(t)=R\dot{\rho}_R(t)R^\dag+R[i D(t,0),\rho_R(t)]R^\dag
.
\eeq
By plugging Eqs.~(\ref{eq: decomposition jump oper}) and (\ref{eq:
dens matr in R representation2}) in Eqs.~(\ref{eq:MEred}) and
(\ref{eq: corr functA}), recalling the action (\ref{eq:intertwin})
of $R(t,0)$ and considering that the eigenvalues of the Hamiltonian
are time independent (Lamb shifts and decay rates will be time
independent too), we obtain the following master equation:
\begin{equation}\label{eq: ME for System}
\dot{\rho}_R(t)=-i[H_S(0),\rho_R(t)]-i[D(t,0),\rho_R(t)]+\lambda^2
\Gamma[\rho_R(t)]
\end{equation}
where
\begin{eqnarray}
\Gamma[\rho_R(t)]&=& \sum_{\alpha,\beta=0,1,\pm}
f_{\alpha\beta}(t)
\left(i\Delta_{\alpha\beta}[L_{\alpha\beta}L_{\alpha\beta}^\dagger,\rho_R(t)]\right.\nonumber\\
&-&\left.\frac{\Gamma_{\alpha\beta}}{2}
\left(\{L_{\alpha\beta}L_{\alpha\beta}^\dagger,\rho_R(t)\}
-2L_{\alpha\beta}^\dagger\rho_R(t)L_{\alpha\beta}\right)\right) . \nonumber\\
\end{eqnarray}
The Lindblad operators read
\beq\label{eq: Lindblad operators}
L_{\alpha\beta}=|D_{\alpha}(0)\rangle\langle D_{\beta}(0)| ,
\eeq
while $f_{\alpha \beta}=f_{\beta \alpha}$,  with
\beq \left.\begin{array}{ccc}
{f_{\alpha0}=\frac{\cos^2{\varphi}}{2},}&
{f_{\alpha1}=\frac{\sin^2{\varphi}\cos^2{\vartheta}}{2},}
&{f_{\alpha\pm}=\frac{\sin^2{\varphi}\sin^2{\vartheta}}{4}}\\
\end{array}\right.
\eeq
and $f_{00}=f_{11}=f_{01}=0$.

In the case of a thermal bath the Lamb shifts and the decay rates
read
\barr
\Delta_{\alpha\beta} & = & P\int_{-\infty}^{\infty} d\omega
\frac{\xi_{\rm{th}}(\omega)}{\omega-\epsilon_{\alpha}+\epsilon_{\beta}},
\nonumber \\
\Gamma_{\alpha\beta}&=&
2\pi\xi_{\rm{th}}(\epsilon_{\alpha}-\epsilon_{\beta})
\label{eq: Lamb shift and diss rates}
\earr where $P$ denotes the principal value,
\barr
\xi_{\rm{th}}(\omega)&=&\xi(\omega)[n_B(\omega)+1] +
\xi(-\omega)n_B(-\omega)\nonumber\\
&=& \frac{1}{2}\xi(\omega)
\left[\coth\frac{\beta\omega}{2}+1\right]\nonumber\\
& & + \frac{1}{2}\xi(-\omega)
\left[\coth\frac{\beta\omega}{2}-1\right]
\label{eq: thermal spectral density}
\earr
is the thermal spectral density, $\xi(\omega)$ being the bare
spectral density of the noise [$\xi(\omega)=~0\,\,
\text{for}\,\,\omega<0$)]
\beq\label{eq: bare spectral density}
\xi(\omega)=\sum_k{\gamma_k}^2\,\delta(\omega_k-\omega)
\eeq
and $n_B(\omega)=1/(\rm{exp} (\beta\omega) -1)$ the mean number of
bosons at frequency $\omega$.

Notice that a non vanishing temperature entails an effective
modification of the form factors of the interaction, making them
in general unbounded from below \cite{tasaki}. Moreover, note that
when $\epsilon_{\alpha}=\epsilon_{\beta}$, Eqs.~(\ref{eq: Lamb
shift and diss rates}) particularize to
\beq
\Delta_{\alpha\alpha}=\int_0^\infty d\omega
\frac{\xi(\omega)}{\omega},
\quad
\Gamma_{\alpha\alpha}=2\pi \frac{\xi'(0^+)}{\beta}
\eeq
and the Lamb shift $\Delta_{\alpha\alpha}$ is temperature
independent.

\section{Evolution in presence of noise: optimal working point}
\label{sec: results}

Equation~(\ref{eq: ME for System}) was numerically integrated
along the loop in Fig.~\ref{fig: NotPath} when the three arcs are
covered at a constant angular speed. We set
$\Gamma_{+0}=\Gamma_{+1}=\Gamma_{0-}=\Gamma_{1-}=1.1\,
\Omega,\, \Gamma_{0+}=\Gamma_{-0}=\Gamma_{1+}=\Gamma_{-1}=0.8\,
\Omega,\, \Gamma_{++}=\Gamma_{--}=1\, \Omega,\,
\Gamma_{+-}=1.2\, \Omega,\,
\Gamma_{-+}=0.7\, \Omega,\,
\Delta_{+0}=\Delta_{+1}=\Delta_{0-}=\Delta_{1-}=-1.1\, \Omega,\,
\Delta_{0+}=\Delta_{-0}=\Delta_{1+}=\Delta_{-1}=0.8\, \Omega,\,
\Delta_{++}=\Delta_{--}=1\, \Omega,\,
\Delta_{+-}=-1.2\, \Omega,$ and
$\Delta_{-+}=0.7\, \Omega $. These values of the Lamb shifts and
decay rates are physically meaningful and have been chosen,
somewhat arbitrarily, for illustrative purposes. We will consider
later the realistic case of an Ohmic bath at different
temperatures and will directly derive the values of all the
constants from Eq.~(\ref{eq: Lamb shift and diss rates}).

In Fig.~\ref{fig: confront} we show the behavior of the fidelity
for three different initial states. In each graph, from top to
bottom, the dissipation constant increases from $\lambda^2=0$ to
0.05.

\begin{figure}
\includegraphics[width=0.95\columnwidth]{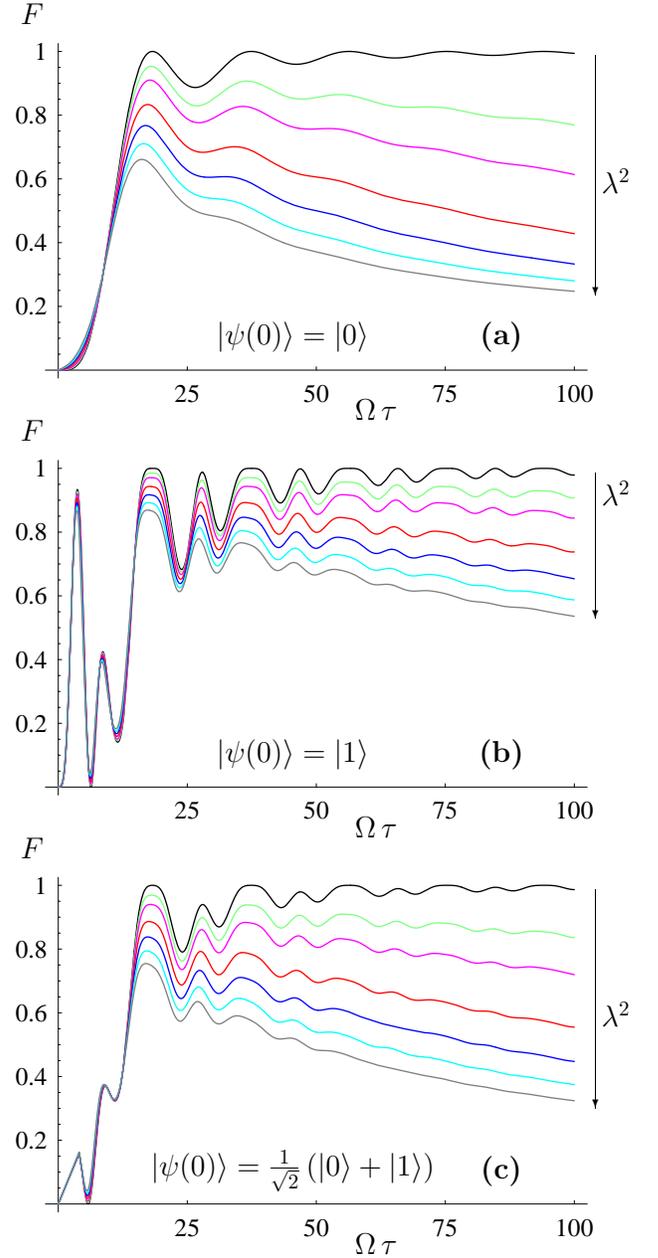}
\caption{(Color online) Fidelity $F$ versus cyclic time $\Omega \tau$
for three different initial states, $\Omega$ being the energy gap
between the bright and dark states and $\tau$ the time needed to
cover the loop shown in Fig.~\ref{fig: NotPath} (NOT gate). (a)
``up'' state; (b) ``down'' state; (c) completely symmetric state.
In each graph the dissipation constant $\lambda^2$, defined in
Eq.~(\ref{eq: ME for System}), increases from top to bottom:
$\lambda^2=0$ (noiseless case), 0.005, 0.01, 0.02, 0.03, 0.04 and
0.05.}
\label{fig: confront}
\end{figure}
In the noiseless case (upper line in the three plots) the fidelity
tends to 1 when $\Omega\tau \to \infty$ (adiabatic limit). This
asymptotic value is not reached monotonically: there are some
oscillations, with maxima at $F=1$ in the noiseless case. This is
the case discussed in Section \ref{sec: Free Ideal Evolution}: the
NOT transformation is perfect, even though one is far from the
adiabatic regime, at the time values given by (\ref{eq: optimal
times for NOT}).

Notice that although the oscillations and the general behavior of
$F$ depend on the initial state, the overall trend is of general
validity and only depends on the path in the parameter space and
the speed at which it is covered. Interestingly, the position of
these maxima is only weakly dependent on noise: the horizontal
shifts of the peaks are almost irrelevant. (This is due to the
small influence of the Lamb shifts.) The mean fidelity (averaged
over a set of input states uniformly distributed over the Bloch
sphere) is displayed in Fig.\ \ref{fig:media}. One observes all
the interesting features discussed before.

\begin{figure}
\includegraphics[width=0.95\columnwidth]{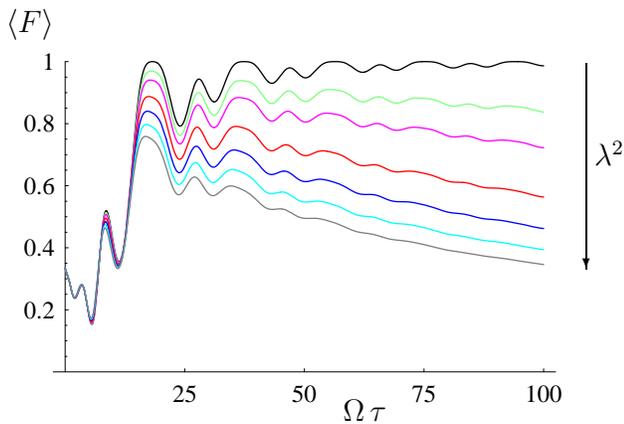}
\caption{(Color online) Mean fidelity $\langle F \rangle$ versus cyclic time $\Omega
\tau$. All parameters are identical to those of Fig.\ \ref{fig:
confront}.}
\label{fig:media}
\end{figure}

\begin{figure}
\includegraphics[width=0.95\columnwidth]{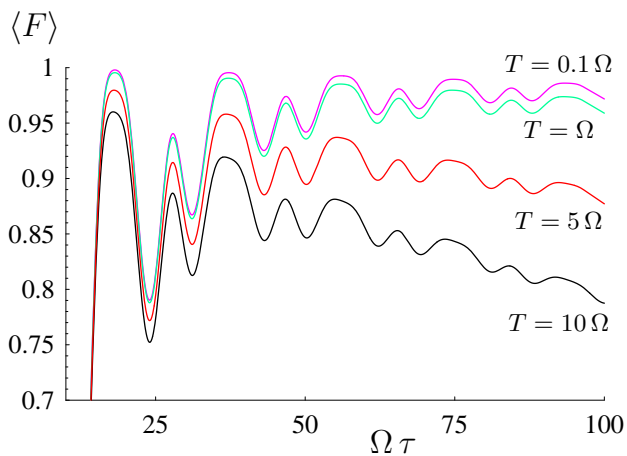}
\caption{(Color online) Mean fidelity $\langle F \rangle$ versus cyclic time $\Omega
\tau$, evaluated for an Ohmic spectral density at different temperatures
$T=\beta^{-1}$, with $\lambda^2=0.01$. Notice the different scale
on the vertical axis, as compared to the previous figures.}
\label{fig: mediaOhmicSD}
\end{figure}

For all the above reasons, we can estimate the value of the
optimal gate-operation point for experimental applications as that
point corresponding to the first peak. An important remark is
however in order: in the non adiabatic regime, the gate is no
longer purely geometrical. Both dynamical and geometrical effects
contribute to the transformation and cannot be easily separated,
because in general the generators do not commute. In principle it
would be possible to extract the geometrical contribution, but one
would not gain any additional information, useful for experimental
purposes.

In general, in the presence of noise, the fidelity decreases as
the time needed for the transformation increases. This can make it
difficult to obtain a pure geometrical transformation (because of
the necessary adiabatic condition). This problem could be slightly
reduced by choosing a large energy gap between the degenerate
computational space and the bright states. However, it seems more
convenient to exploit the presence of the peaks and partially
neglect this physical request. As a matter of fact, the fidelity
decrease due to the noise is very small in the non adiabatic
regime. This makes the NOT gate feasible by using a fine tuning of
the total operation time. The best performance is obtained for an
\emph{optimal operation time}
\beq
\tau^*  =\tau^*_{k=1}= \frac{3 \pi }{2 \Omega}\sqrt{15}
\label{eq: optimal time}
\eeq
corresponding to the first peak of the fidelity.

On the other hand, some interesting features of the fidelity do
depend on the initial states and are apparent in Fig.\ \ref{fig:
confront}. Some states are less robust than others; this is due to
the fact that, during the evolution, the population transfer
between the levels depends on the initial state: thus, the longer
the population ``lives'' in a level which is subject to noise, the
less efficient is the transformation.  A critical point is
obviously the total amount of noise. In our simulations we have
considered a noise strength $\lambda^2$ ranging from $0.5\%$ to
$5\%$. A realistic physical estimate yields a noise not exceeding
$0.5\%$. In this regime the fidelity at the optimal point reaches
values greater than $0.9$ for all the states considered. From this
result it is clear that we can exploit the optimal times for
realizing the NOT transformation with a relatively high fidelity
even in absence of additional control.

These general conclusions can be corroborated by considering a
particular spectral density for the environment: Figure~\ref{fig:
mediaOhmicSD} display the behavior of the mean fidelity for an
Ohmic spectral density
\beq
\xi(\omega)=\kappa \,
\omega \,\rm{exp}(-\omega/\omega_c)
\eeq
at different temperatures. We used Eq.~(\ref{eq: Lamb shift and
diss rates}) and set $\kappa=1/100$, $\omega_c=100 \,\Omega$ and
$T=1/\beta=\Omega/10,\Omega,5\Omega,10\Omega$, corresponding to a
thermal bath at low, intermediate, high and very high
temperatures, respectively. Notice that the mean fidelity is
always higher than $80\%$ in the whole range of times considered
(up to 100 $\Omega \tau$). In particular, at the optimal time,
fidelity decreases only up to a few percent, even in the very high
temperature case.

As already emphasized at the end of Sec.\ \ref{sec: Free Ideal
Evolution}, these results can be extended to more general loops,
yielding optimal times like in Eq.~(\ref{eq: generalization of
optimal times}). This can be of interest for experimental
applications.

\section{Conclusions}
\label{sec: conclusions}
We studied some aspects of holonomic quantum computation by
focusing our attention on a particular physical system, shown in
Fig.\ \ref{fig: levels}, undergoing a loop in the parameter space
that, in the adiabatic limit, yields a NOT gate. We derived a
general expression for the evolution operator, without taking the
adiabatic limit. After specializing this formula to a specific
loop, an exact expression for the propagator was obtained.  In
order to gain more physical insight we considered the fidelity of
the transformation compared to the adiabatic one. We found that
there exist some values for the duration of the evolution for
which the fidelity is equal to $1$, even though one is far from
the adiabatic regime. The presence of these peaks is important for
experimental applications: if the total operation time can be fine
tuned to the first peak, one can realize a transformation which is
the most robust against noise. In particular, we considered
several initial conditions and analyzed the effects of quantum
noise on the evolution of the system (if the rate of change of the
system Hamiltonian is much smaller than the typical timescales of
the thermal bath). As an example, we considered a particular
noise, inducing transition outside the computational space, and
obtained the relevant master equation. Its numerical solution
yielded information on the behavior of the fidelity and showed how
important the optimal points is: actually, it enables one to
obtain (without external additional control) high values of the
fidelity before the system suffers much from the detrimental
consequences of the noise. It will be important to understand
whether and how this feature can be of help when one tries to
control the system in order to reduce decoherence and dissipation.

We conclude by emphasizing that the strategy suggested in this
article in order to minimize the effects of decoherence, being
based on the determination of a (finite, non adiabatic) optimized
operation time, is somewhat different from the other strategies
suggested so far for suppressing decoherence. Clearly, these
observations can be of interest for experimental applications. One
should try and understand whether the fidelity revivals, and
consequently optimal operation times, exist also for two-qubit
gates.

\acknowledgments This work was supported by the European Community
through contracts EuroSQIP, IST-SQUBIT2, IST-TOPQIP. One of us
(G.F.) thanks Scuola Normale Superiore in Pisa for their kind
hospitality and support.


\appendix
\section{}\label{sec: Appendix A}
We give here the exact analytical expressions of the operators
that describe the evolution of the system in the loop shown in
Fig. \ref{fig: NotPath}. A straightforward but lengthy calculation
yields (in the basis $\{\ket{D_i(0)}\}_{i=0,1,+,- }$)
\beq
U_{\pi/2}(\Omega\tau)=U_{3}(\Omega\tau_3)\,U_{2}(\Omega\tau_2)\,U_{1}(\Omega\tau_1)
\label{eq:UU}
\eeq
with (we set $\tau_1=\tau_2=\tau_3=\tau/3$ for simplicity)
\begin{widetext}
\beq
    U_1(\Omega\tau/3)=\left(\begin{array}{cccc}
    1&0&0&0\\
    0&{\frac{\pi \,\sin \alpha }{2\,\alpha }}&{\cos \alpha
    -\frac{ i \Omega \tau \sin\alpha}{3\sqrt{2}\alpha}}&
    {\frac{3\alpha\cos\alpha+i\Omega\tau\sin\alpha}{3\sqrt{2}\alpha}}\\
    0&{-\frac{{\sqrt{2}}\,
      \left(  2\,i   \,\Omega \,
         \tau  + 3\,\pi \,\cos \alpha  \right) }
      {\beta } }&{\frac{3\,\pi  +  2\,i   \,
     \Omega \,\tau \,\cos \alpha  +
    6\alpha \,\sin \alpha }{\beta }}&{\frac{-3\,\pi  - 2\,i  \,
     \Omega \,\tau \,\cos \alpha  +
    6\alpha \,\sin \alpha }{\beta }}\\
    0&{\frac{{\sqrt{2}}\,\left( 2\,i
         \,\Omega \,\tau  -
      3\,\pi \,\cos \alpha  \right) }{\beta^* }}&{\frac{-3\,\pi  + 2\,i  \,
     \Omega \,\tau \,\cos \alpha  +
    6\alpha \,\sin \alpha }{\beta^* }}&{\frac{3\,\pi  - 2\,i   \,
     \Omega \,\tau \,\cos \alpha  +
    6\alpha \,\sin \alpha }{\beta^* }}\\
    \end{array}\right),
\eeq
\end{widetext}

\begin{widetext}
\beq
    U_2(\Omega\tau/3)=\left(\begin{array}{cccc}
    {\frac{\pi \,\sin \alpha }{2\,\alpha }}&{\cos\alpha}&{\frac{-i \,\Omega \,\tau \,
    \sin \alpha }{{3\sqrt{2}}\,\alpha }}&{\frac{i \,\Omega \,\tau \,
    \sin \alpha }{{3\sqrt{2}}\,\alpha }}\\
    {\frac{-\left( 4\,{\Omega }^2\,{\tau }^2 +
      9\,{\pi }^2\,\cos \alpha  \right) }{36\,
    {\alpha }^2}}&{\frac{\pi \,\sin \alpha }{2\,\alpha }}&{\frac{i\,\pi \,\Omega \,\tau \,
    \left( -1 + \cos \alpha  \right) }{
    {6\sqrt{2}}\,{\alpha }^2}}&{\frac{-i \,\pi \,\Omega \,\tau \,
    \left( -1 + \cos \alpha  \right) }{
    {6\sqrt{2}}\,{\alpha }^2}}\\
    {\frac{i\,\pi \,\Omega \,\tau \,
    \left( -1 + \cos \alpha  \right) }{
    {6\sqrt{2}}\,{\alpha }^2}}&{\frac{-i \,\Omega \,\tau \,
    \sin \alpha }{{3\sqrt{2}}\,\alpha }}&{\frac{9\,{\pi }^2 + 2\,{\Omega }^2\,{\tau }^2 +
    2\,{\Omega }^2\,{\tau }^2\,\cos \alpha }
    {36\,{\alpha }^2}}&{-\frac{{\Omega }^2\,{\tau }^2\,
      \left( -1 + \cos \alpha  \right)   }
    {18\,{\alpha }^2}}\\
    {\frac{-i\,\pi \,\Omega \,\tau \,
    \left( -1 + \cos \alpha  \right) }{
    {6\sqrt{2}}\,{\alpha }^2}}&{\frac{i \,\Omega \,\tau \,
    \sin \alpha }{{3\sqrt{2}}\,\alpha }}&{-\frac{{\Omega }^2\,{\tau }^2\,
      \left( -1 + \cos \alpha  \right)   }
    {18\,{\alpha }^2}}&{\frac{9\,{\pi }^2 + 2\,{\Omega }^2\,{\tau }^2 +
    2\,{\Omega }^2\,{\tau }^2\,\cos \alpha }
    {36\,{\alpha }^2}}\\
    \end{array}\right),
\eeq
\end{widetext}

\begin{widetext}
\beq
    U_3(\Omega\tau/3)=\left(\begin{array}{cccc}
    {\frac{\pi \,\sin \alpha }{2\,\alpha }}&0&{-\frac{{\sqrt{2}}\,
      \left( 2\,i \,\Omega \,
         \tau  + 3\,\pi \,\cos \alpha  \right) }
      {\beta }}&{\frac{{\sqrt{2}}\,\left( 2\,i
         \,\Omega \,\tau  -
      3\,\pi \,\cos \alpha  \right) }{\beta^* }}\\
    0&1&0&0\\
    {\frac{3\,\alpha \,\cos \alpha  -
    i \,\Omega \,\tau \,\sin \alpha }{3\,
    {\sqrt{2}}\,\alpha }}&0&{\frac{3\,\pi  + 2\,i  \,
     \Omega \,\tau \,\cos \alpha  +
    6\,\alpha \,\sin \alpha }{\beta }}&{\frac{-3\,\pi  + 2\,i   \,
     \Omega \,\tau \,\cos \alpha  +
    6\,\alpha \,\sin \alpha }{\beta^* }}\\
    {\frac{3\,\alpha \,\cos \alpha  +
    i \,\Omega \,\tau \,\sin \alpha }{3\,
    {\sqrt{2}}\,\alpha }}&0&{\frac{-3\,\pi  - 2\,i  \,
     \Omega \,\tau \,\cos \alpha +
    6\,\alpha \,\sin \alpha }{\beta }}&{\frac{3\,\pi  - 2\,i  \,
     \Omega \,\tau \,\cos \alpha  +
    6\,\alpha \,\sin \alpha }{\beta^* }}\\
    \end{array}\right),
\eeq
\end{widetext}
where $\alpha=\sqrt{9\,{\pi }^2 +
       4\,{\Omega }^2\,{\tau }^2}/6$ and $\beta=6\,\pi  + 4\,i  \,\Omega \,
   \tau$.
These are the operators appearing in Eq.~(\ref{eq: U analytic in
adiab lim}) [with $\tau_i=\tau/3,\,\,i=1,2,3$ and the angular
velocities in Eq.\ (\ref{eq: def fidelity}) kept constant].

\section{}\label{sec: Appendix B}

We derive here the master equation (\ref{eq:MEred}), by focusing
on the physical aspects of the proof, in the context of quantum
computation with holonomic gates.

Let us start by considering the Liouville operator (\ref{eq:
liouvillianA}) acting on density matrices $\rho$. Consider a
density operator $\varrho(t)$ describing the system and the
environment. Its evolution is governed by the von
Neumann-Liouville equation (\ref{eq: def eq liouvilleA}), with the
initial condition (\ref{eq: factorization+bath equilA}). In the
interaction picture engendered by the free Liouvillian
$\mathcal{L}_0(t)$, \beq\label{eq: def inter}
\Phi_0(t,0)=\textbf{T}\exp\left\{\int_{0}^{t}\mathcal{L}_0(s)ds\right\},
\eeq the density matrix takes the form \beq\label{eq: def rho
inter} \varrho_I(t)={\Phi_0}^{-1}(t,0)\varrho(t) \eeq and
Eq.~(\ref{eq: def eq liouvilleA}) reads \beq\label{eq: eq
liouville inter}
\dot{\varrho}_I(t)=\mathcal{L}_{SB}(t,0)\varrho_I(t), \eeq where
\beq
\mathcal{L}_{SB}(t,0)={\Phi_0}^{-1}(t,0)\mathcal{L}_{SB}\Phi_0(t,0).
\eeq By formally integrating (\ref{eq: eq liouville inter}) we get
\beq\label{eq: def evol rho inter}
\varrho_I(t)=\textbf{T}\exp\left\{\int_{0}^{t}\mathcal{L}_{SB}(s,0)ds\right\}
\varrho(0), \eeq whose expansion up to second order in the
interaction Liouvillian $\mathcal{L}_{SB}$ reads
\begin{eqnarray}
\varrho_I(t)& = &\left[1+\int_{0}^{t}ds\mathcal{L}_{SB}(s,0)\right.\nonumber\\
& & +\left.\int_{0}^{t}ds\int_{0}^{s}du\mathcal{L}_{SB}(s,0)
\mathcal{L}_{SB}(u,0)\right]\varrho(0) .\nonumber\\
\label{eq: sec order exp}
\end{eqnarray}
The reduced density operator of the system is obtained by tracing
(\ref{eq: sec order exp}) over the bath:
\begin{eqnarray}
\sigma_I(t)&=&\tr_B\left\{\varrho_I(t)\right\}\nonumber\\
&=&\sigma(0)+\int_{0}^{t}ds\tr_B\left\{\mathcal{L}_{SB}(s,0)\varrho(0)\right\}\nonumber\\
& & +\int_{0}^{t}ds\int_{0}^{s}du\tr_B\left\{\mathcal{L}_{SB}(s,0)
\mathcal{L}_{SB}(u,0)\varrho(0)\right\} .\nonumber\\
\label{eq: evol reduced rho}
\end{eqnarray}
The second term in this expansion vanishes for a bath at equilibrium
(\ref{eq: factorization+bath equilA}). Therefore Eq.~(\ref{eq: evol
reduced rho}) can be written as
\begin{eqnarray}
\sigma_I(t)&=&\sigma(0)+\nonumber\\
&+&\int_{0}^{t}ds{\Phi_S}^{-1}(s,0)\left[\int_{0}^{s}du\,
K(s,u)\right]{\Phi_S}(s,0)\sigma(0) ,\nonumber\\
\label{eq: evol reduced rho2}
\end{eqnarray}
where
\beq\label{eq: def phi0}
\Phi_S(t,0)=\textbf{T}\exp\left\{\int_{0}^{t}\mathcal{L}_S(s)ds\right\},
\eeq
\beq\label{eq: kernel ME}
K(s,u)=
\tr_B\left\{\mathcal{L}_{SB}\tilde{\mathcal{L}}_{SB}(s,u)\sigma_B\right\}\eeq
and
\beq
\tilde{\mathcal{L}}_{SB}(s,u)={\Phi_0}(s,u)\mathcal{L}_{SB}{\Phi_0}^{-1}(s,u).
\eeq
At this point one assumes that the typical timescale of the
evolution is much slower than the timescales characterizing the
bath and derives a master equation in the Markov approximation. We
make a further hypothesis, justified by the physical nature of the
process we intend to study. In our system (\ref{eq: Schroedinger
equation})-(\ref{eq: decomp H}) there is another timescale,
related to the rate of change of the system Hamiltonian: we assume
that this is the slowest timescale of our problem, due to the
adiabaticity of the evolution and then, compared to the bath
correlation time, the evolution of $\mathcal{L}_S$ is always
``adiabatic." This is assured by the condition (\ref{eq: condition
on LSA}).
This condition allows us to write
\begin{eqnarray}
\Phi_S(s,u)&=&\textbf{T}\exp\left\{\int_{u}^{s}\mathcal{L}_S(t)dt\right\}\nonumber\\
&\simeq&\textbf{T}\exp\left\{\int_{u}^{s}\mathcal{L}_S(s)dt\right\}
=\exp\{\mathcal{L}_S(s)(s-u)\}\nonumber\\
\label{eq: condition on phiS}
\end{eqnarray}
when $|s-u|<\tau_c$. Moreover the bath correlation function
rapidly vanishes for times larger than the correlation time
$\tau_c$
\beq\label{eq: condition on K}
K(s,u)\simeq 0, \qquad  \text{for} \quad |s-u|>\tau_c .
\eeq
Using Eqs.~(\ref{eq: kernel ME}) and (\ref{eq: condition on phiS})
we obtain
\begin{eqnarray}
K(s,u)&\simeq&\tr_B\big\{\mathcal{L}_{SB}\exp[\mathcal{L}_0(s)(s-u)]\nonumber\\
&\times&\mathcal{L}_{SB}\exp[-\mathcal{L}_0(s)(s-u)]\sigma_B\big\}
\label{eq: approx for K}
\end{eqnarray}
and thus, in the secular approximation, forced by  Davies's
projection, $\sum_{\omega}Q_{\omega}(s)[\cdots]Q_{\omega}(s)$,
\begin{eqnarray}
\Gamma(s)&=& \sum_{\omega}Q_{\omega}(s)
\left[\int_0^{s}du K(s,u)\right]Q_{\omega}(s) \nonumber\\
&\simeq& \sum_{\omega}Q_{\omega}(s)\int_{-\infty}^{0}du
\tr_B\left\{\mathcal{L}_{SB}\exp[-\mathcal{L}_0(s)\,u]\right.\nonumber\\
& & \times \left.\mathcal{L}_{SB}
\exp[\mathcal{L}_0(s)\,u]\sigma_B\right\}Q_{\omega}(s) ,
\label{eq: corr funct}
\end{eqnarray}
where $Q_{\omega}(s)$, given by (\ref{eq: Liouvillian projectorsA}),
are the instantaneous eigenprojections of $\mathcal{L}_S$.
Therefore, Eq.~(\ref{eq: evol reduced rho2}) takes the form
\beq\label{eq: evol reduced rho3}
\sigma_I(t)=\sigma(0)+\int_{0}^{t}ds {\Phi_S}^{-1}(s,0)\Gamma(s)
{\Phi_S}(s,0)\sigma(0) ,
\eeq
which yields the differential equation
\beq\label{eq: ME sigmaI}
\dot{\sigma}_I(t)={\Phi_S}^{-1}(t,0)\Gamma(t){\Phi_S}(t,0)\sigma_I(t).
\eeq
Going back to the Schr\"{o}dinger picture,
\beq\label{eq: int pict sigma}
\sigma(t)={\Phi_S}(t,0)\sigma_I(t) ,
\eeq
we finally get
\begin{eqnarray}\label{eq: ME sigma}
\dot{\sigma}(t)
=[\mathcal{L}_{S}(t)+\Gamma(t)]\sigma(t) .
\end{eqnarray}
This is the master equation (\ref{eq:MEred}): it is the same
equation one would obtain by considering $\mathcal{L}_S(t)$
``frozen" at time $t$, by applying the standard Markov
approximation and evaluating the decay rates and the frequency
shifts at the instantaneous eigenfrequencies
$\omega(t)=\epsilon_m(t)-\epsilon_n(t)$ of the system Liouvillian.
This supports our physical intuition of a system Hamiltonian
adiabatically changing in a faster environment.


\end{document}